\documentclass[journal=jacsat,manuscript=article]{achemso}

\usepackage[version=3]{mhchem} % Formula subscripts using \ce{}
\usepackage{graphicx}% Include figure files
\usepackage{dcolumn}% Align table columns on decimal point
\usepackage{bm}% bold math
\usepackage{xspace}
\usepackage{physics}
\usepackage{svg}
\usepackage{float}
\usepackage{tabto}
\usepackage{array}
\usepackage{makecell}
\usepackage{nicefrac}
\usepackage[utf8]{inputenc}
\usepackage[T1]{fontenc}
\usepackage{mathptmx}
\usepackage{amsmath,amssymb,mathtools}

\newcommand{\nbse} {NbSe$_2$\xspace}
\newcommand{\mos}{MoS$_2$\xspace}
\newcommand{\wse}{WSe$_2$\xspace}

\newcommand{\beginsupportinginfo}{%
        \setcounter{figure}{0}
        \renewcommand{\thefigure}{S\arabic{figure}}%
        \setcounter{equation}{0}
        \renewcommand{\theequation}{S\arabic{equation}}%
     }

\author{T. R. Devidas}
\author{Itai Keren}
\author{Hadar Steinberg}
\email{hadar@phys.huji.ac.il}

\affiliation{The Racah Institute of Physics, The Hebrew University, Jerusalem 91904, Israel}

\title{Spectroscopy of NbSe$_2$ using Energy-Tunable Defect-Embedded Quantum Dots}

\keywords{American Chemical Society, \LaTeX}

\begin{document}

\begin{abstract}
Quantum dots have sharply defined energy levels, which can be used for high resolution energy spectroscopy when integrated in tunneling circuitry.
Here we report dot-assisted spectroscopy measurements of the superconductor \nbse, using a van der Waals device consisting of a vertical stack of graphene-\mos-\nbse. 
The \mos tunnel barriers host naturally occurring defects which function as quantum dots, allowing transport via resonant tunneling.
The dot energies are tuned by an electric field exerted by a back-gate, which penetrates the graphene source electrode. 
Scanning the dot potential across the superconductor Fermi energy, we reproduce the \nbse density of states which exhibits a well-resolved two-gap spectrum. 
Surprisingly, we find that the dot-assisted current is dominated by the lower energy feature of the two \nbse gaps, possibly due to a selection rule which favors coupling between the dots and the orbitals which exhibit this gap.
\end{abstract}

\section{Introduction}
The density of states (DOS) of a superconductor $N_{S}(\epsilon)$ and its evolution across critical parameters (temperature $T$, magnetic field $B$) is well mapped out in tunneling spectroscopy measurements~\cite{Giaever1960}~. The emergence of the van der Waals (vdW) family of materials and the related stacking techniques has paved the way for the fabrication of planar tunneling devices of very high quality, leading to high resolution spectroscopy of layered superconductors~\cite{Dvir2018,Dvir2018a,Zalic2019}~. The tunnel barriers in these vdW tunneling devices consist of the insulator hexagonal Boron Nitride (hBN) and of transition metal dichalcogenide (TMD) semiconductors such as \mos or \wse~\cite{Dvir2018,Dvir2018a,Zalic2019}~. 
These materials provide a robust and reliable platform for tunnel barriers amenable to exfoliation. Barrier thickness can be pre-determined with a single atomic layer precision, which enables versatile control over the tunnel junction resistance. The transparency of the tunnel junction~\cite{Britnell2012}~, which is related to the barrier resistance, plays a decisive factor in precise spectroscopy of the tunneling DOS since attaining the lowest sub-gap conductance requires suppression of Cooper pair tunneling~\cite{Greibe2011,Island2016}~.
\\

Van der Waals barriers host atomic point defects~\cite{Chakraborty2015,He2015,Tran2017,Hong2015,Vancso2016}~ associated with localized energy states addressable in electrical transport measurements~\cite{Chandni2015,Greenaway2018,Papadopoulos2020, Keren2020}~. A recent study from our group by Dvir et al.~\cite{Dvir2019}~, showed that when these barrier-bound defects are placed in proximity to a vdW superconductor, the heterostructure emulates the quantum dot (QD) - superconductor (SC) systems often studied in nanowires (NW)~\cite{DeFranceschi2010,Deacon2010,Lee2014,Scherubl2020}~. The ground-state of a QD-SC system is defined by the relative coupling strength of the QD with the normal lead ($\Gamma_{N}$), the superconducting lead ($\Gamma_{SC}$) and the superconducting energy gap value $\Delta$. In the regime where $\Gamma_{SC} > \Delta$, the defect couples strongly to the SC, giving rise to Andreev Bound State (ABS) features at $\varepsilon < \Delta$. In the weakly coupled regime, where $\Gamma_{SC,N} << \Delta$, QD-SC transport is dominated by single electron resonant tunneling processes. At this limit, the dot-assisted transport, through the sharply defined energy level of the QD, can serve as a sensitive spectrometer~\cite{Deng2016,Junger2019}~. 
\\

In the present study we report on tunneling measurements of a graphene-\mos-\nbse heterojunction device where the \mos barrier harbors embedded QDs. The potential of these dots is tuned by application of voltage to the back-gate, placed under the graphene source electrode. Differential conductance maps, traced vs. source-drain ($V_{SD}$) and gate ($V_{BG}$) voltages, take the form of a truncated coulomb diamond (CD)~\cite{Junger2019,Wang2018}~, where the superconducting energy gap suppresses transport at zero bias. \nbse is a two-gap superconductor~\cite{Boaknin2004,Zehetmayer2010}~, which exhibits two spectral signatures in tunneling measurements, $\Delta_1\sim$ 1.2 mV and $\Delta_2\sim$ 0.6 mV~\cite{Noat2015,Dvir2018}~.  
Our dot-assisted tunneling measurements reveal two distinct conductance onsets, which are clearly related to these two gaps. 
Interestingly, the dot appears to be selectively more sensitive to the low energy gap - in contrast to standard tunneling measurements~\cite{Dvir2018}~, which are more sensitive to the high energy gaps. We suggest that such coupling could either be a consequence of overlap in momentum states, or of the particular orbital of the QD and its coupling to the \nbse Se orbitals.

\section{Results and discussion}
Figure 1(a) shows the schematic~\cite{Momma2011}~ of the heterostructure  measured in the current experiment. We measured three devices of the same geometry, all of which exhibit QD resonant tunneling. The main report focuses on Device 1 which exhibits features from an isolated dot superimposed on a clear \nbse tunneling spectrum. The hatched region indicates the active tunneling junction area (J.A.) in Device 1, which is $\sim 16$ $\mu m^{2}$. Data from the two other devices are reported in the Supporting Information.
The optical image of Device 1 with the boundaries of the flakes denoted with coloured outlines and labelled correspondingly is shown in Figure 1(b). Graphene was exfoliated onto a standard $p$-doped silicon substrate with 285 nm oxide. The tunnel barrier \mos flake and superconductor \nbse flake (HQ Graphene, Netherlands) were exfoliated under an inert Argon atmosphere in a glovebox and transferred sequentially onto the graphene by a dry-stamp technique using PDMS~\cite{Castellanos-Gomez2014}~. Ti/Au electrodes were evaporated on graphene and \nbse after processing through standard e-beam lithography. For electrodes on \nbse, a pre-evaporation ion milling step using Argon ions is carried out for obtaining oxide-free low resistance contacts. 
\\

\begin{figure}[H]
\centering
\includegraphics[width = 350 pt]{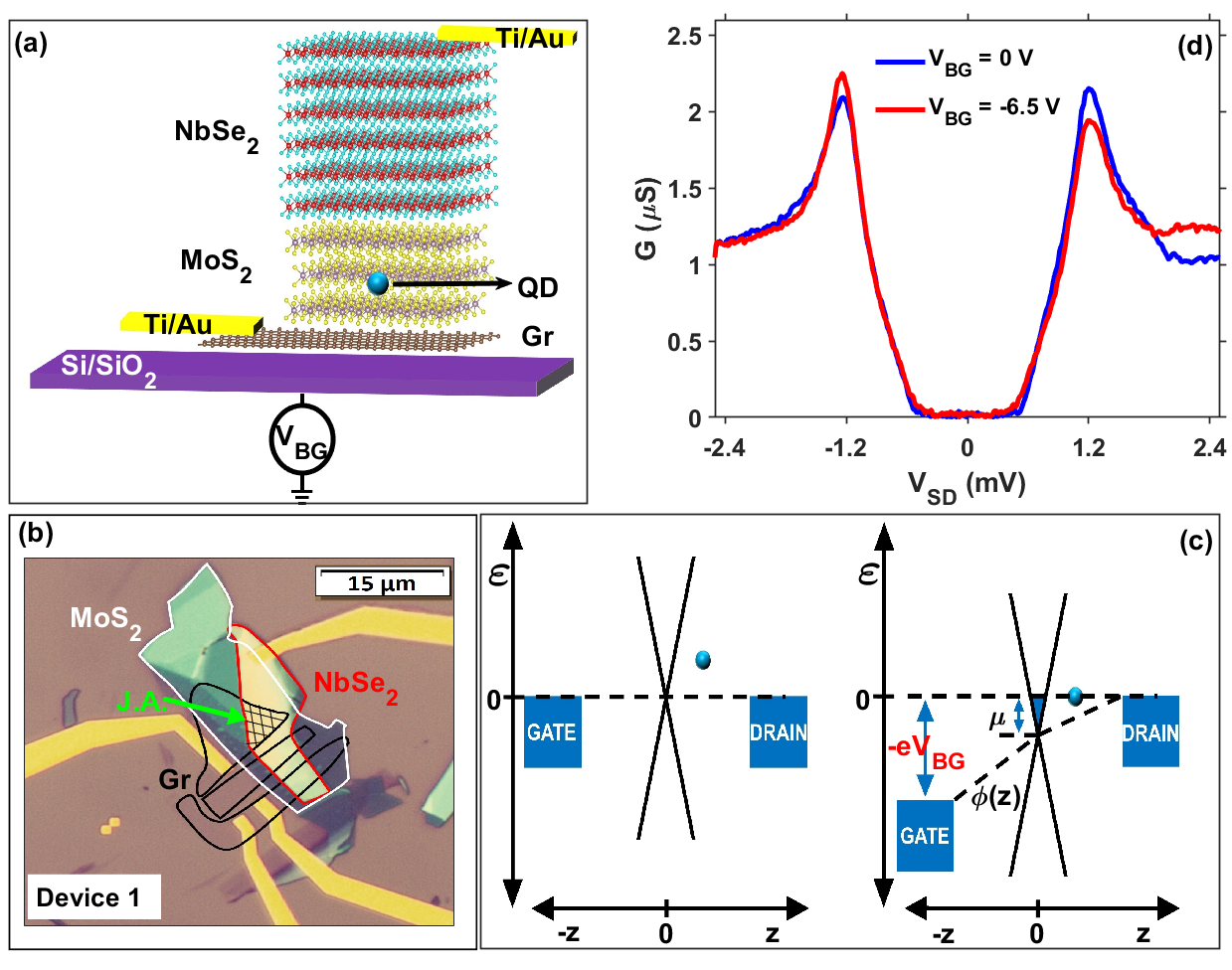}
\caption{(a) Schematic of the device measured in the current work; (b) Optical image of Device 1 with boundaries and labels pointing to the different materials. The hatched region J.A., is the active tunneling junction area; 
(c) Schematic explaining the Gate-induced electric field penetration: As graphene is charged, it's dispersion shifts downward, resulting in a sloped potential and shift to the energy of the barrier embedded QD; (d) Differential conductance scans of the device at two different $V_{BG}$ values. The contribution from the QD results in gate-tunable changes to the spectrum.}
\end{figure}
\vspace{10 pt}

The schematics in Figure 1(c) show how a graphene electrode permits the penetration of electric field exerted by an external gate~\cite{Greenaway2018,Papadopoulos2020,Keren2020}~, by tracking the evolution of the dot potential $\mu_{dot}$ with the application of back gate voltage $V_{BG}$.
The left panel depicts the potential map assuming the electrochemical potential $\phi(z)$ is constant at $V_{BG}$ = 0 with graphene at zero density. Here, $z$ is the vertical coordinate, set to 0 at the graphene layer. Without external fields, the dot potential $\mu_{dot}$ can attain a broad range of initial values, which depend on its position and chemical identification. 
Application of $V_{BG} > 0$  shifts $\phi(z)$ downwards and charges the graphene, increasing its chemical potential to a positive value, $\mu_{Gr}>0$. 
For a given graphene voltage ($V_{Gr} = 0$ in the figure), the change in density results in $\phi(0)=-\mu_{Gr}$. The resulting tilt of $\phi(z)$ at $z > 0$ corresponds to the penetration of the gate-induced electric field through the graphene layer, causing a downward shift in $\mu_{dot}$. The change in $\mu_{dot}$ translates to an effective dot-gate capacitance, which depends on graphene DOS~\cite{Keren2020}~. The dot-assisted transport can then be analyzed using the standard dot transport model , assuming separate capacitance values between the dot and (i) the source ($C_S$), (ii) the drain ($C_D$) and (iii) the gate ($C_G$)~\cite{Keren2020,Wang2018}~.
\\

Figure 1(d) shows the differential conductance $G = dI/dV_{SD}$ vs. bias $V_{SD}$ measurement at a sample temperature of $T$ = 30 mK, under two different applied backgate voltages ($V_{BG}$). The spectra exhibit strong quasiparticle peaks at $V_{SD} \sim$ 1.2 mV and a well-developed `hard-gap'~\cite{Dvir2018}~, the ratio between conductance at zero bias ($G_0$) and conductance outside the gap ($G_N$) i.e. $G_0/G_N \approx 1/450$.
Additionally a gate dependent variation between the two spectra is observed. We associate this additional contribution to conductance from a QD embedded in the barrier \mos and define it as $G_d$.
The total measured differential conductance $G$ can then be written as the sum of a direct tunneling contribution $G_t$ and the dot assisted tunneling contribution $G_d$. 

\begin{equation} \label{gen_tunnel_conductance_eq.}
    G = \frac{dI}{dV_{SD}} = G_{t} +G_{d}
\end{equation}

Here, $G_t$ is directly proportional to the convolution of DOS of the source (graphene) and the drain (superconductor)~\cite{Tinkham2004,Bardeen1961}~. 
\\

Figure 2(a) shows the evolution of $G$ as a function of $V_{BG}-V_{SD}$ at zero magnetic field ($B = $ 0 T). Riding on a nearly constant spectral map, the additional conductance from the QD enabled channel ($G_d$) is seen to form a truncated coulomb diamond (CD)~\cite{Junger2019,Wang2018,Keren2020}~. 
We average the measured spectra over the entire $V_{BG}$ range so as to obtain an approximation to $G_t$, via smearing the dot contribution. This $G_t$ is then subtracted from measured $G$ to obtain the QD contribution to the spectra ($G_d$). This $G_d$ over the entire $V_{BG}-V_{SD}$ range is plotted in Figure 2(b).
\\

In Figure 2(c) we track the QD transport features over a wide range in $V_{BG}$ and $V_{SD}$. We note that this scan was taken in a different cooldown, and hence the QD resonant feature was shifted to a more negative $V_{BG}$. We find that even when scanning $V_{BG}$ over a range of $\pm 90$ V, no evidence of a successive Coulomb diamond appears~\cite{Keren2020}~,
setting a lower limit for the charging energy of $E_C > 50$ meV. Such high charging energies, associated with a very small volume of the QD, are consistent with previous reports~\cite{Keren2020, Greenaway2018, Papadopoulos2020}~.
We also notice that at these high biases, the QD transport features exhibit finite curvature - due to the change in QD-Gate capacitance associated with a change in the graphene DOS~\cite{Keren2020}~.

\begin{figure}[H]
\centering
\includegraphics[width = 375 pt]{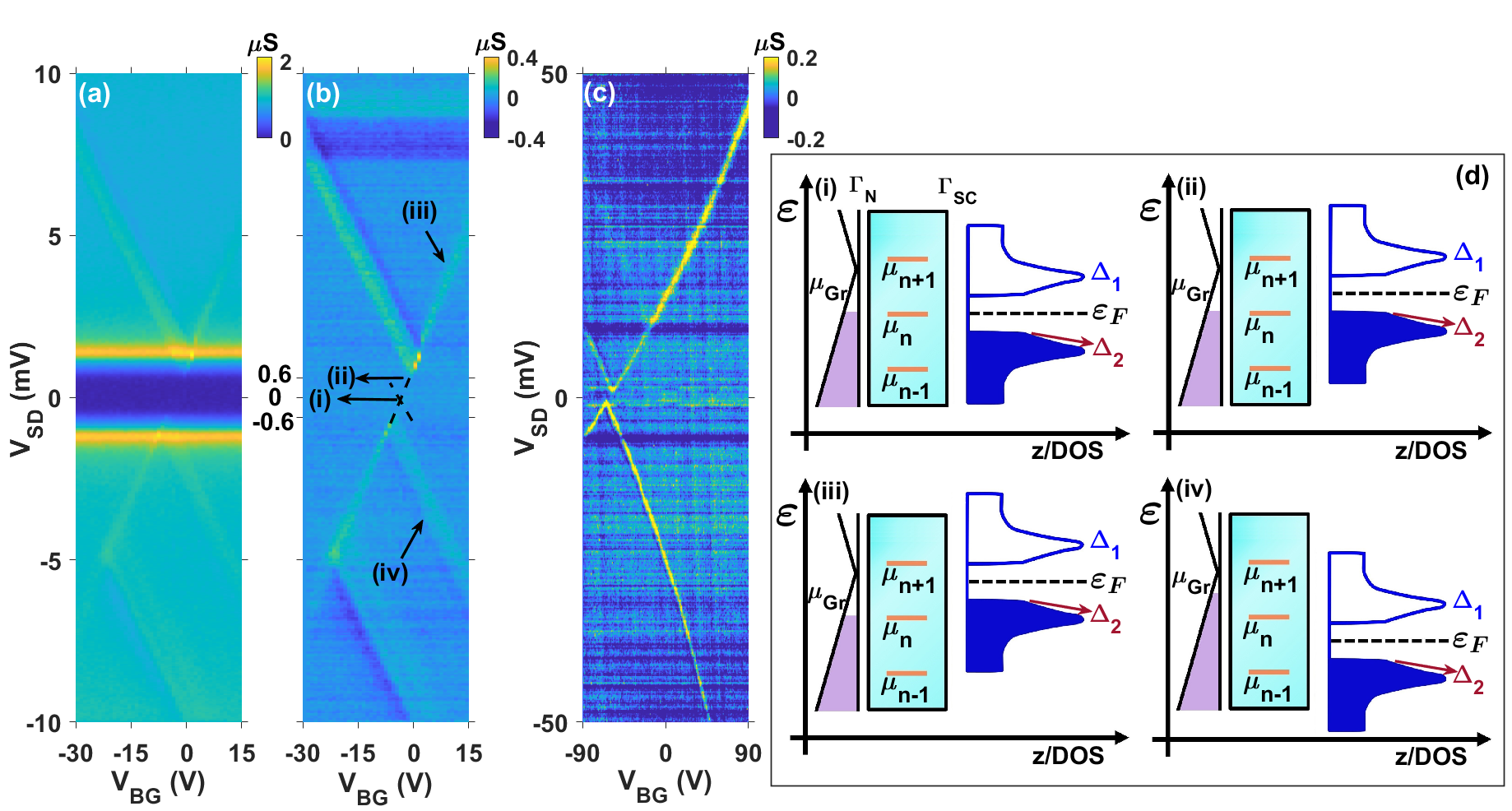}
\caption{(a) Differential conductance $G$ map as a function $V_{BG}-V_{SD}$; (b) Dot assisted differential conductance map $G_d$ after subtraction of a constant contribution from the measured spectrum $G$ shown in (a);(c) $G_d$ map of the same QD in a second cooldown, over a greater $V_{BG}-V_{SD}$ range ; (d) Schematics depicting the various Graphene-QD-SC tunneling conditions (i - iv), explained in the text}%
\end{figure}
\vspace{10 pt}

The schematics presented in Figure 2(d) demonstrate how the QD serves as a spectrometer. $\mu_{n-1,n,n+1}$ define the chemical potentials of the QD at different charging states. Labels (i - iv) in Figure 2(b) correspond to the illustrations in Figure 2(d). The intersection point of the two black dashed lines forming an "x" in Figure 2(b), labelled (i), corresponds to the the condition $V_{SD}$ = 0 mV. This is the degeneracy condition where the QD's chemical potential, $\mu_{dot} = \mu_n$, is in resonance with both $\mu_{Gr}$ and $\mu_{SC}$, and transport is suppressed due to the SC energy gap.
As one moves along the positive $V_{BG}-V_{SD}$ diagonal, resonant tunneling via the QD starts at $V_{SD}$ = 0.6 mV (label (ii)), where the dot is resonant with both $\mu_{Gr}$  and the onset of finite DOS at the SC. Further along the same slope (label(iii)), $\mu_{dot}$ remains in resonance with $\mu_{Gr}$. This is termed as the "Source resonance" condition. Label (iv) is the drain resonance condition, where ${\mu_{dot}}$ is in resonance with onset of finite DOS at the superconductor.
Along the source resonant diagonal $\mu_{dot}$ always stays at fixed bias $V_{SD}$. As we show below, spectral information is gained by tracking the evolution of $G_d$ parallel to this trajectory.

\begin{figure}[H]
\centering
\includegraphics[width = 300 pt]{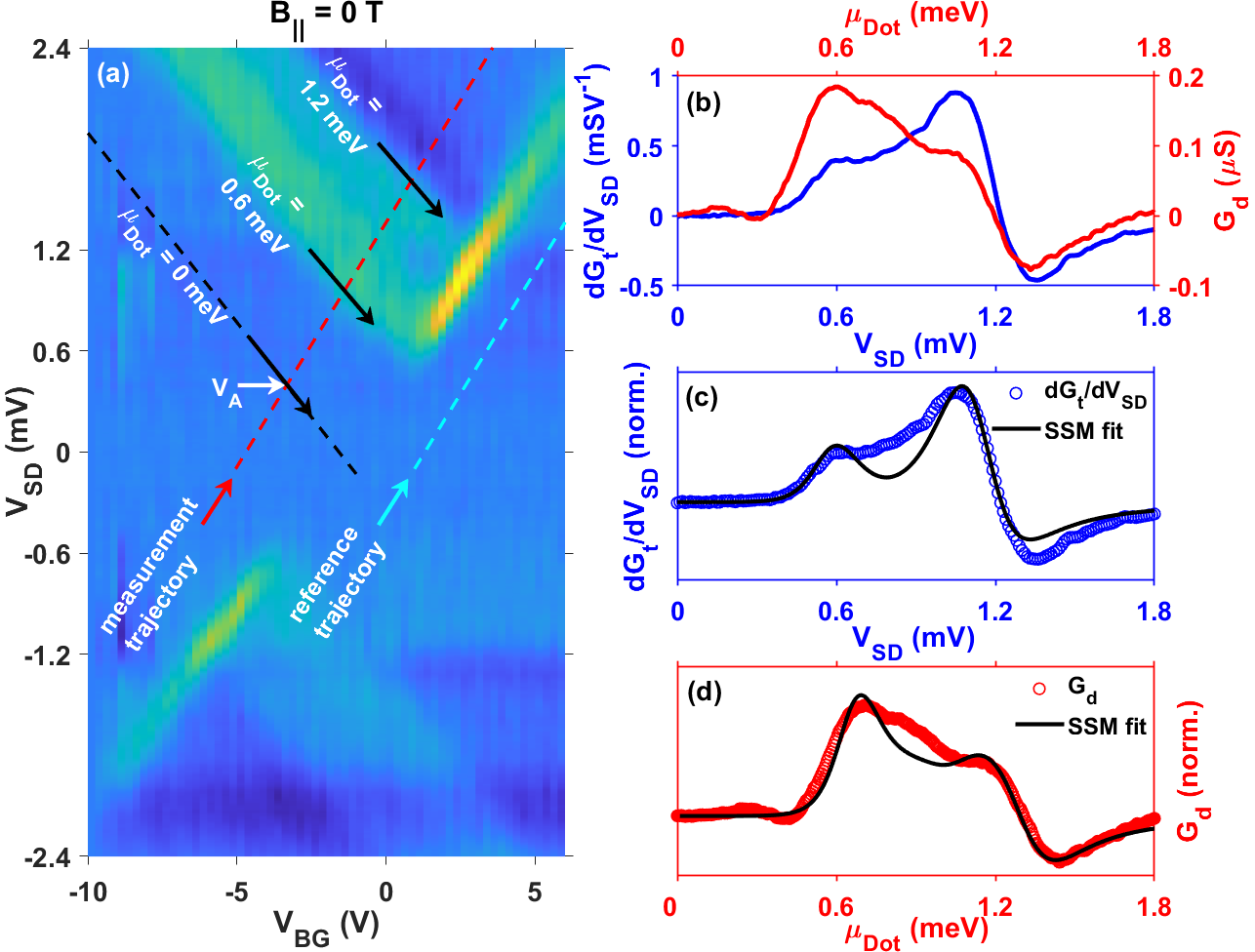}
\caption{(a) High resolution map of $G_d$ as function of $V_{BG}-V_{SD}$. The black dashed lines and the black arrows indicate diagonals along which the chemical potential of the QD is a constant. The measurement trajectory (red dashed line) and the reference trajectory (cyan dashed line), indicate the two measured line scans along $V_{BG}-V_{SD}$ diagonals parallel to the source resonance condition; (b) Comparison between the derivative of tunneling $dG_t/dV_{SD}$  vs. $V_{SD}$ (left y-axis, blue line) and dot assisted tunneling $G_{d}$ vs. dot energy $\mu_{dot}$ (right y-axis, red line); (c) $dG_t/dV_{SD}$ and (d) $G_d$ vs. fit to the SSM model (black line). The fit parameters are tabulated in Table 1.}%
\end{figure}
\vspace{10 pt}

QD-assisted spectroscopy of \nbse is demonstrated in Figure 3. In panel (a) we present a high-resolution subtracted conductance map i.e. $G_d(V_{BG},V_{SD})$ focusing on the cross feature. 
The map clearly shows the superconducting gap, extending to $\pm 0.6$ mV. Conductance onset corresponding to drain-resonance (Figure 2(d) panel (iv)) appears along two parallel features, representing the two gaps of \nbse. 
Such data can be translated into spectral measurements by tracking the trajectory marked by a red dashed line by stepping simultaneously $V_{BG}$ and $V_{SD}$ . From this measured data, $G_d$ is obtained by subtracting a data measured along a reference trajectory, indicated by the cyan dashed line which is parallel to the red dashed line. This can be translated to $G_d(\mu_{dot})$ by noting that along this trajectory,  $\mu_{dot} = V_{SD}-V_A$, where $V_A$ is the point where QD is resonant with the drain Fermi energy along this particular trajectory (marked in panel (a)).  
$G_d(\mu_{dot})$ is shown on the right $y$-axis (red) in panel (b). It exhibits a strong onset at $\mu_{dot} = 0.6$ meV, followed by a plateau like feature at $\mu_{dot} = 1.2$ meV.
\\

To interpret this result we recall from Equation~\eqref{gen_tunnel_conductance_eq.} that the total conductance $G$ is a sum of the direct tunneling $G_t$ and QD assisted tunneling $G_d$. The DOS of the QD can be approximated to that of a Dirac-Delta distribution $\delta(\varepsilon-\mu_{dot})$. 
We start with the expression for dot-assisted tunneling current: 

\begin{equation} \label{Eq_Landauer_Dot}
I(V_{SD}) = \frac{G_{N-N}}{e}\int_{-\infty}^\infty \delta(\varepsilon-\mu_{dot}) \frac{N_S(\varepsilon)}{N_S(0)}\left[f(\varepsilon)-f(\varepsilon+eV_{SD}) \right] d\varepsilon
\end{equation}

where $G_{N-N} = A |T|^2 N_{Gr}(0)N_S(0)$. $N_{Gr}$ and $N_S$ are the DOS of the graphene and superconductor respectively, $A$ is a constant of proportionality, $T$ is the tunneling matrix element. Taking the derivative with respect to $V_{SD}$, we obtain the expression for $G_d$ (see Supporting Information):

\begin{equation} \label{dot_cond_eq}
\begin{gathered}
\begin{multlined}
G_d = -\frac{G_{N-N}}{N_S(0)} \bigg \{ \alpha_s \frac{\partial N_S(\mu_{dot})}{\partial\varepsilon}[f(\mu_{dot})-f(\mu_{dot}+eV_{SD})]+ \ldots \\
\alpha_s N_S(\mu_{dot}) \bigg[\frac{\partial f(\mu_{dot})}{\partial \varepsilon}-\frac{\partial f(\mu_{dot}+eV_{SD})}{\partial \varepsilon} \bigg] 
+ N_S(\mu_{dot})\delta(\mu_{dot}+eV_{SD}) \bigg \}
\end{multlined}
\end{gathered}
\end{equation}

We find that $G_d$ is proportional to the derivative of the DOS of the superconductor, with $\alpha_s$ a capacitive leverage factor - via the first term on the R.H.S. of equation \eqref{dot_cond_eq}. The other terms appear only when the QD is resonant with the source or drain electro-chemical potentials.
\\

We thus compare $G_d(\mu_{dot})$ to the derivative of the tunneling conductance $dG_t/dV_{SD}$. The latter exhibits the typical structure, associated with the two band nature~\cite{Dvir2018}~ of superconductivity in bulk \nbse, with spectral features at $V_{SD} \sim$ 0.6 mV and 1.2 mV.  
$G_d(\mu_{dot})$ exhibits the same features, albeit with a difference observed in their relative intensity: The dot assisted tunneling data shows stronger coupling with the lower energy feature $\Delta_2$, against $G_t$'s stronger coupling with the higher energy feature $\Delta_1$. 
\\

Two-band superconductivity in \nbse has been described based on the McMillan model~\cite{McMillan1968}~ where superconductivity in a band possessing a larger gap (Nb derived bands at $\Gamma$ and $K$ pockets in the Brillouin zone) induces superconductivity in a second band, possessing a smaller gap (Se derived bands at the $\Gamma$ pockets)~\cite{Kiss2007}~. 
How these two bands manifest in the tunneling spectra, is determined by the relative selectivity of the tunneling electron to each of them.
Tuning such selectivity can be done geometrically - for example, in scanning tunneling spectroscopy measurements by Noat et al,~\cite{Noat2015}~ selectivity was controlled by engaging the tip to different crystallographic orientations. 
It is clear from Figure 3(b), that tunneling electrons whose origin is associated with a defect in \mos, could play a role in determining band selectivity.
\\

Following the previous report~\cite{Dvir2018}~ from our group on vdW tunneling in \nbse, we analyse the measured spectra, both $dG_t/dV_{SD}$ and $G_d$ using the Suhl, Schopohl, McMillan [SSM] model~\cite{Suhl1959,Schopohl1977,McMillan1968}~. The coupled equations, whose solutions give the superconducting band gaps $\Delta_i$, i = 1,2, are 

\begin{equation}
    \Delta_i(\varepsilon) = \frac{\Delta_i^0 + \left(\nicefrac{\gamma_{ij} \Delta_j(\varepsilon)}{\sqrt{\Delta_{j}^2(\varepsilon)-\varepsilon^2}}\right)}{1+\left(\nicefrac{\gamma_{ij}}{\sqrt{\Delta_{i}^2(\varepsilon)-\varepsilon^2}}\right)}
\end{equation}

where $\Delta_i^0$ is the intrinsic gap arising from electron-phonon coupling in each band and $\gamma_{ij}$ is the rate of quasiparticle scattering between bands $i,j$. 

The total DOS in each band is given by

\begin{equation}
    N_i(\varepsilon) =  T_i N_{i}(\varepsilon_{F}) \frac{1}{2 \pi} \int d\theta R \frac{\abs{E}}{\sqrt{( 1 + \alpha cos\theta)} \Delta_{i}^2(\varepsilon) - \varepsilon^2}
\end{equation}

where $T_i N_{i}(\epsilon_{F})$ is the effective DOS including the tunnel selectivity term $T(i)$ and $\alpha$ is the band anisotropy parameter. A least square routine is employed to fit the curve to the measured $dG_t/dV_{SD}$ (Figure 3(c)) and $G_d$ (Figure 3(d)) respectively. The fit parameters are tabulated in Table I.

\vspace{10pt}
\begin{table}[H]
    \centering
    \label{t_sim}
    \begin{tabular}{|c|c|c|c|c|c|c|c|c|}
    \hline
    \thead{Spectrum} & $\Delta_1^0$ & $\Delta_2^0$ & $\gamma_1$ & $\gamma_2$ & $N_1/N_2$ & $T$ & $\alpha$ \\
    \hspace{0.1 cm} & (meV) & (meV) & (meV) & (meV) & & (K) & \\[1pt]
    \hline
    \hline 
    \thead{$dG_t/dV_{SD}$} & 1.17 & 0.1 & 0.3 & 1.3 & 14.28 & 0.05 & 0\\[1pt]
    \hline
    \thead{$G_d$} & 1.25 & 0.1 & 0.3 & 1.8 & 2 & 0.05 & 0.1\\[1pt]
    \hline
    \end{tabular}
    \caption{SSM fit parameters to the spectra in Figure 3(c) and 3(d)}
    \label{tab:my_label}
\end{table}
\vspace{10pt}

The values of the fit parameters for the spectrum $dG_t/dV_{SD}$ lie appreciably close to the values reported for vdW tunneling into bulk \nbse by Dvir et al~\cite{Dvir2018}~. 
We note that in the SSM model, the value of $\Delta_1^0$ obtained from the fit is highly reliable. However, $\Delta_2^0$ cannot be determined precisely, as its value depends on the choice of the scattering rates. We find that values of $\Delta_2^0$ between 0 meV to 0.3 meV, in combination with $\gamma_2$, all yield reasonable fits to the data.
The results of the SSM fits indicate that QD-assisted tunneling differs from direct tunneling in two main parameters : (a) tunneling selectivity, manifest in the ratio of the DOS of the two bands $N_{1}/N_{2}$ - where we find the relative contribution from the inner band $N_2$ being $\approx 7$ times greater, and (b) the band anisotropy parameter $\alpha$. 
\\

It is not clear how dot-assisted tunneling is sensitive to band anisotropy. It is possible, however, to suggest a number of possible origins to the tunneling selectivity of the dot towards the lower energy feature. A foremost contender is the fact that the barrier embedded defect-dot could be fortuitously located directly above a Selenium atom in the \nbse. Such a position-induced selectivity has been earlier observed in STM by Guillamon et al.~\cite{Guillamon2008}~, when the tunneling tip was located directly above the Selenium atom.
Alternatively, the selectivity could be influenced by wave-function overlap - noting that the Selenium $4 p_z$-derived bands extend further away from the \nbse surface, in comparison to the Niobium $4d$-derived bands~~\cite{Noat2010,Rahn2012}~. 
A third option could be related to overlap in momentum space. In \mos, the most commonly occurring defect, due to it's lowest energy of formation is the sulphur mono-vacancy ($V_S$). Bound to a localised $V_S$, electron states in a monolayer \mos are known to extend to a typical diameter of 6 \r{A}~\cite{Qiu2013}~. This suggests that coupling from a defect state would favor bands placed within $\delta k = 1$ \r{A}$^{-1}$ from the $\Gamma$ point. These would favor the Se and Nb-derived bands centred at the $\Gamma$ pocket in~\nbse~\cite{Rahn2012}~. 
Finally, we may also consider the orbital symmetry of the $V_S$. If vacancy and chalcogen orbitals share the same symmetry, such similarity would influence the tunnel selectivity. Needless to say, based on a single dot we can not differentiate between these possibilities.
\\

We complete our investigation by presenting measurements of dot-assisted tunneling $V_{BG}-V_{SD}$ maps with applied in-plane magnetic fields, shown in Figure 4. At $B_\parallel = 1$ T (Figure 4(a)), the zero-bias cross feature is observed. This is a consequence of orbital depairing of smaller energy gap $\Delta_2$, associated with the Se-like bands at the $\Gamma$ point. As seen in Dvir et al.~\cite{Dvir2018}~, these carriers are characterized by a greater diffusion coefficient and are prone to depairing by the application of $B_\parallel$. This creates a continuum of sub-gap states, allowing the recovery of the zero-bias cross-over feature (at 1 T), as observed in typical Coulomb blockade devices. 
\\

Further increasing the magnetic field to $B_\parallel = 2$ T (Figure 4(a)), one of the branches of the Coulomb diamonds appears to split, exhibiting a clearly resolved excited level, which is associated with the Zeeman splitting of the QD energy. Figures 4(b) to 4(e) show the $G_d$ map with increasing magnetic fields upto 6 T. Figure 4(f) shows the complete evolution of the Zeeman feature energies (tracked by red arrows in Figures 4(a)-(e)) up to $B_{||} = 6$ T at a constant bias $V_{SD} = 1.2$ mV. The gate-dependent resonant features can be converted to dot energy using the formula $\Delta E = \alpha_{CD}\Delta V_{BG}$, where $\alpha_{CD} \approx 1/8300 $ eV/V is the sum of inverse of the slopes of the two QD resonant diagonals in the coulomb diamond (CD). The Zeeman split features exhibit an energy split $E_Z = \pm g\mu_{B}B$ with Land\'e factor $g \approx 1.7$. Recalling that strongly coupled defect dots exhibit $g$ factors ranging between 1.3 and 2.0~\cite{Dvir2019}~, we suggest that the weakly coupled QD observed here originates from similar type of defect.

\begin{figure}[H]
\centering
\includegraphics[width = 375 pt]{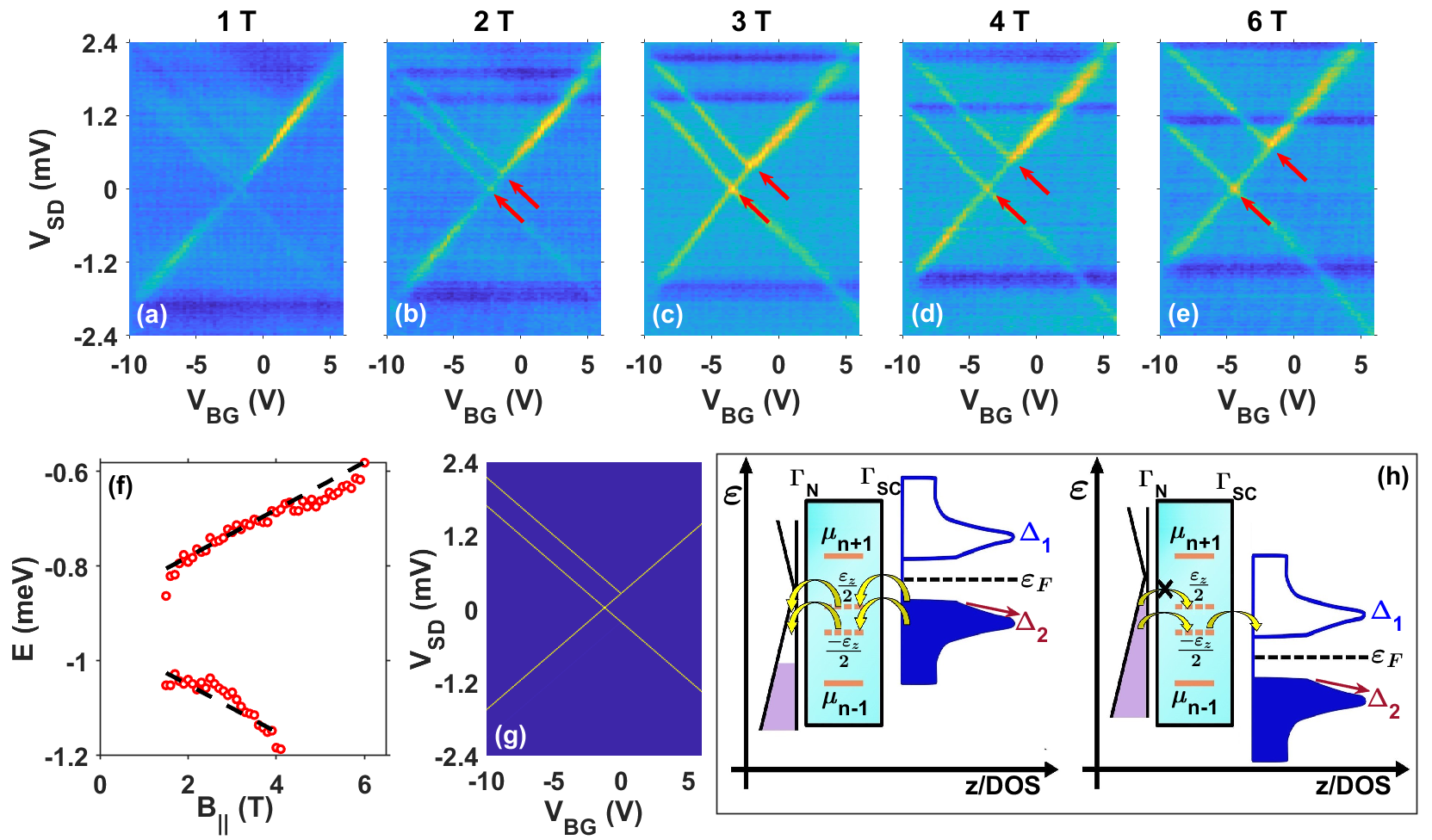}
\caption{(a-e) $G_d(V_{BG},V_{SD})$ maps at $B_{||}$ = 1 T, 2 T, 3 T, 4 T and 6 T; (f) Evolution of Zeeman feature energies, converted from $V_{BG}$ at $V_{SD}$ = 1.2 mV (see text), tracked by the red arrows, with applied magnetic field $B_{||}$; (g) Model for Dot assisted tunneling current for $B_{||} = 2$ T; (h) schematic of the tunneling process in the presence of non-zero magnetic field (B $\geq$ 2 T). An asymmetric coupling of the QD to the graphene and the superconductor, $\Gamma_N >> \Gamma_{SC}$, leads to the observation of Zeeman splitting only on positive bias.}%
\end{figure}
\vspace{10pt}

It is also seen that the Zeeman split feature appears only on one side, the positive bias side of the coulomb diamond. This indicates that the dot is asymmetrically coupled to the graphene and superconductor electrodes. To see this, we solve a rate-equation model for a positive bias large enough to allow transport through both levels - where electrons tunnel from the SC to the spin-split QD, and subsequently to graphene. Dot occupancy is solved using the coupled rate equations for the Zeeman-split levels $\pm \varepsilon_z/2$ whose occupancies are denoted as $n_{u,d}$: 
\begin{equation} \label{rateeq.}
    \frac{dn_{u,d}}{dt} = \Gamma_{SC}(1-n_u-n_d)-\Gamma_{N} n_{u,d} \mp \Gamma_r n_u(1-n_d)
\end{equation} 
Where $\Gamma_r$ is the relaxation rate between the levels. 
Summing the two equations (cancelling out the relaxation term), leads to the steady-state occupancy:
\begin{equation} \label{sspop.}
    (n_u+n_d) = \frac{2\Gamma_{SC}}{2\Gamma_{SC}+\Gamma_N}
\end{equation} 
The current is $j = 2e\Gamma_N\Gamma_{SC}/(2\Gamma_{SC}+\Gamma_N)$. 
An electron occupying any of the spin-split states blockades the other one, as accounted by the first term on the right hand side of Equation~\eqref{rateeq.}.
Thus, a subsequent carrier can only hop on to the dot if both are unoccupied.
At this regime, tunneling through both channels will be possible if $\Gamma_{N} >> \Gamma_{SC}$, ensuring that the levels are unoccupied prior to the subsequent tunneling from the SC to the QD. We thus have $j=2e\Gamma_{SC}$, and the excited state feature will be seen for $B_{||}$ = 2 T in Figure 4(g). 
At the reverse bias the coupling rates in Equation~\eqref{rateeq.} reverse roles, yielding $j = -2e\Gamma_N\Gamma_{SC}/(\Gamma_{SC}+2\Gamma_N)$.
Here, an electron tunneling on to the QD from the graphene side would drain slowly to the SC and thus limit transport to a single channel,  $j=-e\Gamma_{SC}$. Hence, in this bias the excited level will not be seen in transport.
\\

We conclude that we have demonstrated the utility of a barrier-bound defect as a tunable energy quantum dot. QD energy, tuned by gating through a graphene source electrode, is scanned across the spectrum of \nbse, demonstrating sensitivity to its two distinct bands. We find that the QD couples strongly to the lower energy gap, as opposed to extended tunneling electrodes where coupling favors the higher energy gap. We discuss possible origins of this selectivity.

\begin{acknowledgement}
The authors thank E. Rossi for valuable discussions. The authors thank Tom Dvir and Shahar Simon for helping with the SSM model codes. Funding for this work was provided by a European Research Council Starting Grant (No. 637298, TUNNEL), Israel Science Foundation grant 861/19, and BSF grant 2016320.

\end{acknowledgement}
\newpage
%\bibliography{References}
\providecommand{\latin}[1]{#1}
\makeatletter
\providecommand{\doi}
  {\begingroup\let\do\@makeother\dospecials
  \catcode`\{=1 \catcode`\}=2 \doi@aux}
\providecommand{\doi@aux}[1]{\endgroup\texttt{#1}}
\makeatother
\providecommand*\mcitethebibliography{\thebibliography}
\csname @ifundefined\endcsname{endmcitethebibliography}
  {\let\endmcitethebibliography\endthebibliography}{}

\beginsupportinginfo
\newpage
\section{Supporting Information}
\subsection{Dot assisted tunneling data on Device 2}

\begin{figure}[H]
    \centering
    \includegraphics[width = 325 pt]{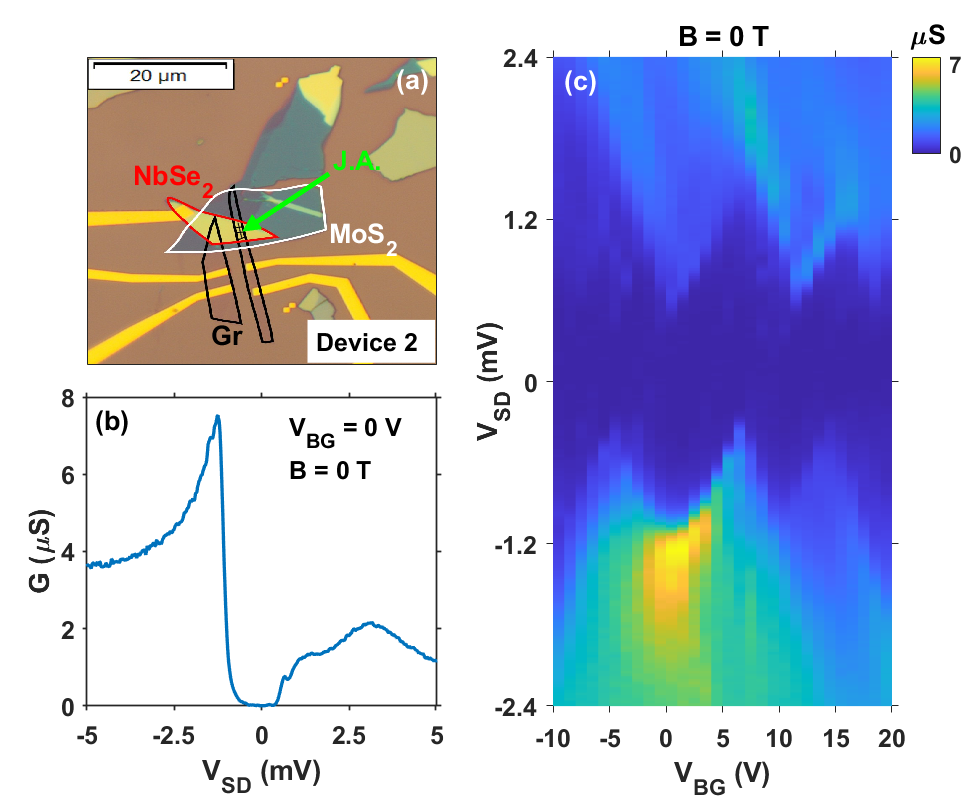}
    \caption{(a) Optical image of Device 2, the hatched region J.A. marks the active junction area; (b) Differential conductance  $G$ as a function of $V_{SD}$; (c) differential conductance map as a function of $V_{BG}-V_{SD}$ at zero applied magnetic field.}
\end{figure}

Figure S1(a) shows the optical image of the device 2, with outlines and corresponding labels given to identify the different flakes. The active tunnel junction area in this device is $\sim 7 \mu m^2$. Figure S1(b) shows the tunneling spectra of the junction as a function of bias $V_{SD}$ at zero gate, $V_{BG}$ = 0 V and zero applied magnetic field. The spectrum exhibits the quasiparticle peak $\sim 1.2$ mV only at negative bias, $V_{SD} < 0$, and behaves oddly on the positive bias side. In the $V_{BG}-V_{SD}$ map, shown in Figure S1(c), multiple dots and the dot assisted tunneling features are observed. On the positive bias $V_{SD} > 0$, the resonant tunneling feature sets in at $V_{SD} \sim 0.6$ mV, similar to the data in Device 1 in the main text. On the negative bias side, resonant tunneling sets in at bias values $V_{SD} < 0.6$ mV.

\newpage
\subsection{Dot assisted tunneling data on Device 3}

\begin{figure}[H]
    \centering
    \includegraphics[width = 325 pt]{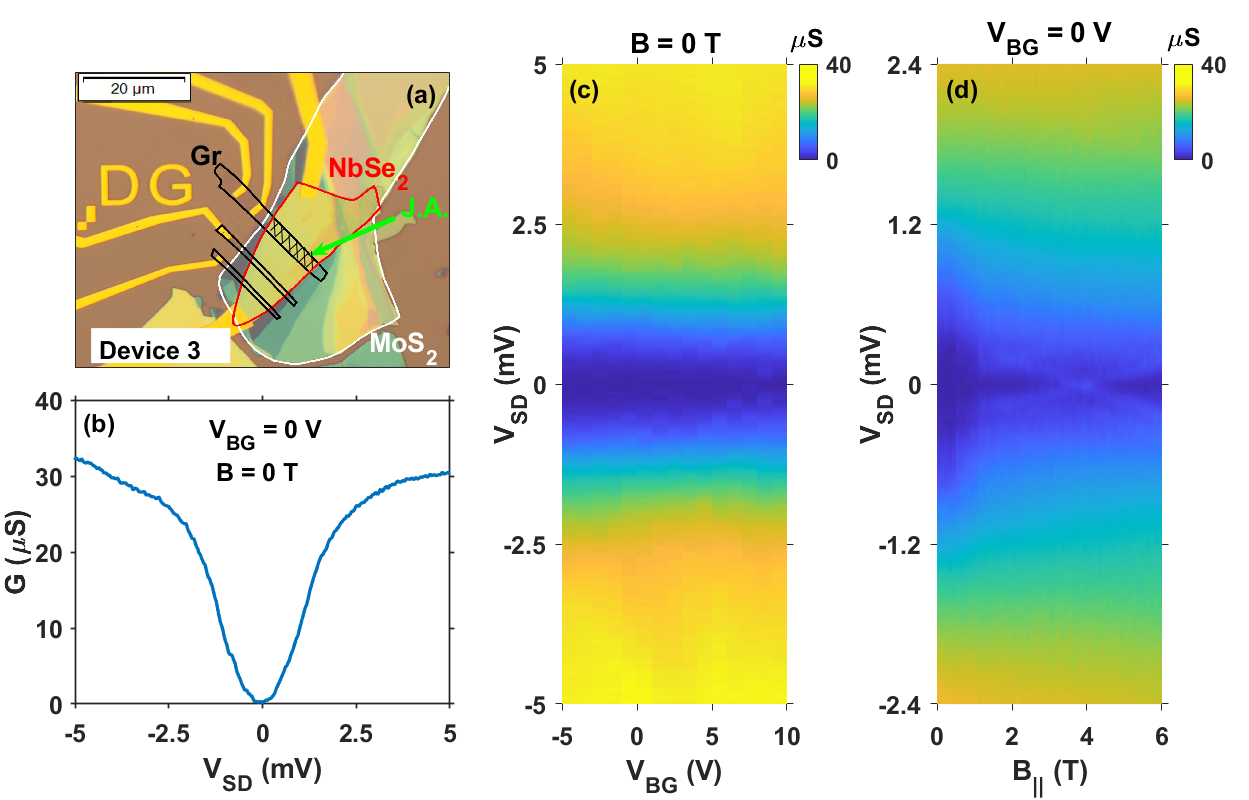}
    \caption{(a) Optical image of Device 3, the hatched region J.A. marks the active junction area; (b) Differential conductance  $G$ as a function of $V_{SD}$; (c) Differential conductance map as a function of $V_{BG}-V_{SD}$ at zero applied magnetic field; (d)Differential conductance map as a function of $V_{SD}$ and applied magnetic field $B_{||}$ at $V_{BG} = 0$.}
\end{figure}
 
Figure S2(a) shows the optical image of Device 3. The active tunnel junction area (J.A.) in this device is $\sim 42 \mu m^2$. Figure S2(b) shows the tunneling spectra as of the junction as a function of bias $V_{SD}$ at zero gate, $V_{BG}$ = 0 V and zero applied magnetic field. The spectrum lacks the quasiparticle features at $V_{SD} \sim $1.1 - 1.2 mV, typical of the high quality tunneling spectrum of $NbSe_2$ reported in the main text. In the $V_{BG}-V_{SD}$ map at zero applied magnetic field, shown in Figure S2(c), features typical of resonant tunneling from Quantum Dots are very weak. In the magnetic field dependence of tunneling spectrum, shown in Figure S2(d), we see a cross-over feature in the spectrum at $B_{||} = 4$ T. However at such high fields of $\sim \frac{1}{4}H_{c2}^{||}$~\cite{suppl_Xi2016}~, the inner second gap $\Delta_2$ undergoes complete depairing and the sub-gap conductance rises subsequently. 

\newpage
\subsection{Tunneling through a quantum dot into a superconductor} 

The net tunneling current through a quantum dot into a superconductor is described by~\cite{suppl_Tinkham2004}~:

\begin{equation} \label{Eq_Landauer_Dot}
I(V_{SD}) = \frac{G_{N-N}}{e}\int_{-\infty}^\infty \delta(\varepsilon-\mu_{dot}) \frac{N_S(\varepsilon)}{N_S(0)}\left[f(\varepsilon)-f(\varepsilon+eV_{SD}) \right] d\varepsilon
\end{equation}
where $G_{N-N} = A |T|^2 N_{Gr}(0)N_S(0)$. $N_{Gr}$, $N_S$ and $\delta(\varepsilon-\mu_{dot})$ are the DOS of graphene, superconductor and QD respectively, $A$ is the constant of proportionality, $T$ is the tunneling matrix element and is assumed to be constant at the low bias voltages applied in the present measurements.

To evaluate the dot assisted differential conductance $G_d$, we take the voltage derivative of the equation~\eqref{Eq_Landauer_Dot}.

The relation between QD energy ($\mu_{dot}$) and the source-drain voltage ($V_{SD}$) is given by $\mu_{dot} = \mu_0-\alpha_s eV_{SD}$, where $\alpha_s$ is a numerical factor depending on the capacitance leverage. 
Taking the voltage derivative of the QD DOS ($\delta$) then leads to -

\begin{align} \label{del_deriv.}
&\frac{\partial}{\partial V_{SD}}\delta(\varepsilon-\mu_{dot}) =  \frac{\partial}{\partial V_{SD}}\delta(\varepsilon-\mu_0+\alpha eV_{SD}) = \alpha_s e\delta'(\varepsilon-\mu_{dot})
\end{align}

Thus the derivative of the tunnelling current is 
\begin{equation} \label{I_tunnel_init1}
\begin{gathered}
\begin{multlined}
G_d = \frac{\partial I(V)}{\partial V_{SD}} = \frac{G_{N-N}}{N_S(0)}\int_{-\infty}^\infty \alpha_s \delta'(\varepsilon-\mu_{dot}) N_S(\varepsilon)\left[f(\varepsilon)-f(\varepsilon+eV_{SD}) \right] d\varepsilon \ldots \\
+\frac{G_{N-N}}{N_S(0)} \int_{-\infty}^\infty \delta(\varepsilon-\mu_{dot}) N_S(\varepsilon)[-f'(\varepsilon+eV_{SD})] d\varepsilon
\end{multlined}
\end{gathered}
\end{equation}

The first integral in equation~$\eqref{I_tunnel_init1}$ is evaluated using the mathematical identity \\
\begin{equation*}
\int_{-\infty}^{\infty} \delta'(x-x_0)f(x)dx = -f'(x_0)
\end{equation*}
leading to

\begin{align} \label{I_tunnel_init1a}
&-\alpha_s\frac{G_{N-N}}{N_S(0)} \left\lbrace \frac{\partial N_S(\mu_{dot})}{\partial\varepsilon}[f(\mu_{dot})-f(\mu_{dot}+eV_{SD})]+N_S(\mu_{dot}) \left[\frac{\partial f(\mu_{dot})}{\partial \varepsilon}-\frac{\partial f(\mu_{dot}+eV_{SD})}{\partial \varepsilon} \right] \right\rbrace
\end{align}

These two terms in equation~$\eqref{I_tunnel_init1a}$ represent the effect of dot energy modulation by source modulation. The $N_S'$ term represents the contribution of dot energy modulation, at the energy window permitted by the source and drain. This measures the derivative of the drain DOS. The second term represents the contribution of the dot resonance with the source and drain. It provides sharp features at the source and drain electrochemical potentials.

The $2^{nd}$ integral in equation~\eqref{I_tunnel_init1} is essentially a condition for coincidence of two delta's. 

\begin{equation} \label{I_tunnel_init1b}
-\frac{G_{N-N}}{N_S(0)}N_S(\mu_{dot})\delta(\mu_{dot}+eV_{SD})
\end{equation}

Physically, this term represents the contribution of the source modulation. When the source potential is resonant with the dot, it probes the drain DOS at the same energy.

Substituting equation~$\eqref{I_tunnel_init1a}$ and equation~$\eqref{I_tunnel_init1b}$ in equation~$\eqref{I_tunnel_init1}$, we get

\begin{equation}
\begin{gathered}
\boxed{\begin{multlined}
G_d = -\frac{G_{N-N}}{N_S(0)} \bigg \{ \alpha_s \frac{\partial N_S(\mu_{dot})}{\partial\varepsilon}[f(\mu_{dot})-f(\mu_{dot}+eV_{SD})]+ \ldots \\
\alpha_s N_S(\mu_{dot}) \bigg[\frac{\partial f(\mu_{dot})}{\partial \varepsilon}-\frac{\partial f(\mu_{dot}+eV_{SD})}{\partial \varepsilon} \bigg] 
+ N_S(\mu_{dot})\delta(\mu_{dot}+eV_{SD}) \bigg \}
\end{multlined}}
\end{gathered}
\end{equation}

\end{document}